# X-ray reflection in Galactic Black Hole Candidates: smeared edge profiles and resonant Auger destruction

R.R. Ross,$^{1,2}$ A.C. Fabian$^2$ and W.N. Brandt$^2$
1. Physics Department, College of the Holy Cross, Worcester, MA 01610, USA (Internet: ross@hcacad.holycross.edu)
2. Institute of Astronomy, Madingley Road, Cambridge CB3 0HA (Internet: acf@ast.cam.ac.uk, wnb@ast.cam.ac.uk)

17 August 1995

**ABSTRACT**
We consider the spectra of Thomson-thick, geometrically-thin accretion discs around Galactic black hole candidates in the reflection model and compute their iron K edges and iron K$\alpha$ lines. We compare the smeared iron K edge profiles that we compute with observation and find them to be a satisfactory description of the data. We find that a combination of Doppler broadening and resonant Auger destruction of line photons can make iron K$\alpha$ lines very difficult to detect in highly ionized inclined discs. We detail the physics of resonant Auger destruction at the level it is currently understood and point out its implications.

## 1 INTRODUCTION

The 1–20 keV spectra of Galactic Black Hole Candidates (BHC) are characterized by independently variable soft blackbody-like components and power laws. Recent work on high quality spectra from EXOSAT and HEAO-1 (Done et al. 1992), Tenma (Kitamoto et al. 1990), Ginga (Ebisawa 1991; Tanaka 1992; Inoue 1993; Ueda, Ebisawa & Done 1994) and BBXRT (Marshall et al. 1993) have revealed spectral features above 6 keV which are probably due to absorption and emission by iron. The strongest feature is an absorption edge occurring above 7 keV which appears smeared relative to a simple photoelectric edge (e.g. Ebisawa 1991). A remarkable aspect of such an edge is that the limits on the equivalent width of any associated emission line are much lower than expected on the basis that about one third of all photons absorbed should result in fluorescent line photons around 6.4 keV (Done et al. 1992; Tanaka 1992; Marshall et al. 1993). There are two ways such a discrepancy might be explained. The first is that the line has been smeared out greatly by Doppler motions. Another is that a large fraction of the fluorescent K$\alpha$ line photons have been resonantly trapped and then destroyed by the Auger effect (sect. II of Ross, Weaver & McCray 1978; Ross & Fabian 1993, hereafter RF). This process requires that the iron atoms being photoionized have between four and ten electrons; i.e. be in the form Fe XVII–Fe XXIII. Comptonization is another mechanism that can destroy the line, but the temperature and optical depth of the Comptonizing medium must be such that it destroys the line but not the edge; i.e. $kT \sim 5$ keV and $\tau_T < 1$. If a corona above the accretion disc provides the illumination then it must have $kT \gg 5$ keV to obtain the observed spectral shape and thus if it destroys the line it will destroy the edge as well. The corona in the model for Cygnus X-1 of Haardt et al. (1993) with $kT \approx 150$ keV and $\tau_T \approx 0.3$ will only destroy the line if the inclination of the disc exceeds $\approx 70$ degrees so the effective optical depth along our line of sight exceeds unity.

The smeared iron K edges of BHC are much deeper than those expected and seen in cold AGN discs (cf. fig. 9 of Matt, Perola & Piro 1991). This would naturally occur in the reflection model if BHC discs were more highly ionized than AGN discs, since stripping of low atomic number elements greatly reduces the photoelectric absorption below 7 keV thereby making the iron edge in the reflection spectrum much more prominent. Smeared iron edges that start around 7.1 keV, the edge energy for Fe I, need not imply that the iron in the disc is in a low ionization state as we shall show below (cf. Marshall et al. 1993).

In this paper we compute spectra of accretion discs around BHC using the Fokker-Planck optically thick radiative transfer code of RF. Our radiative transfer and ionization structure calculations are carried out for fully non-LTE disc plasma (electrons are taken to be in a local Maxwellian distribution due their large elastic scattering cross sections but the radiation field is not Planckian and the atoms do not follow the Saha-Boltzmann law) and incoherent Compton scattering is included. We exhibit the predicted smeared edge profiles from accretion discs and discuss how they can be used to learn about disc properties. We also detail the resonant Auger destruction process and point out its implications. The results presented here, which highlight the edge and line profile in BHC, are complementary to and extend those presented in Matt, Fabian & Ross (1993) which concentrate on the emission line profiles from irradiated discs in AGN.

## 2  THE MODEL AND COMPUTED DISC SPECTRA

The geometry we adopt is that of X-ray reflection and is similar to that assumed in much previous work (e.g. Ebisawa 1991; Done et al. 1992; Marshall et al. 1993; Ueda et al. 1994). Simple fits of the iron absorption edge suggest that any absorbing gas is only moderately ionized and therefore any transmission model involves large amounts of soft X-ray absorption. This is contrary to observation and so simple transmission models can be ruled out. Partial-covering models, in which some of the underlying power-law continuum escapes directly to us, alleviate this problem somewhat and also provide acceptable fits (Marshall et al. 1993).

In reflection models the primary power-law continuum is both directly visible to us and incident on plasma where it is absorbed, scattered and reprocessed to yield a further observed component, the reflection spectrum. The iron edge and any line would be associated with this component. The plasma is usually assumed to be the inner parts of the accretion disc around the black hole.

Our basic model is that of a geometrically thin, $\alpha$-viscosity accretion disc (Shakura & Sunyaev 1973) around a 10 $M_\odot$ Schwarzschild black hole. We consider the disc region between 6 and 26 black hole gravitational radii ($r_g = Gm/c^2$) and consider the radiative transfer in the outer 3–10 Thomson depths (this corresponds to a physical depth of $\sim$ 50 m). We adopt cosmic element abundances (Morrison & McCammon 1983) and energy generation rate proportional to gas density. The resulting equations for the hydrogen number density ($n_H$) and the disc flux ($F_s$), valid for a radiation-pressure-dominated disc, can be found in Ross, Fabian & Mineshige (1992). We take the flux distribution deep in the disc to be a blackbody with temperature $T_s$ so that $B_\nu(T_s) = F_s$ but then solve for the full radiative transfer of the photons in the outer Thomson depths, including incoherent Compton scattering. We take the viscosity parameter $\alpha = 0.1$ and the energy conversion efficiency $\eta = 0.083$ (Shakura & Sunayev 1973).

We consider disc illumination by an optically thin corona in which magnetic dissipation takes place, as is illustrated in fig. 3 of Galeev, Rosner & Vaiana (1979). We assume Compton, bremsstrahlung, and synchrotron processes in the corona cause it to radiate a power-law spectrum with photon index $\Gamma = 1.7$ which isotropically strikes the disc from above and photoionizes its surface layers. We take the coronal illuminating flux at the surface of the disc, $F_h$, to equal $F_s$ (a reasonable assumption for lack of any alternative since the illumination power is ultimately derived via accretion). Details of the treatment of the illumination are given in RF.

Using the code of RF, we compute spectra from 10 irradiated annuli between 6–26$r_g$, chosen so that the flux from each annulus is roughly 10 per cent of the total flux considered. The successful operation of the code requires the plasma to be highly ionized which is why we do not consider radii greater than 26$r_g$ or luminosities less than $0.1L_{\rm Edd}$. Line formation is calculated in an escape probability formalism (Ross 1978), and we assume that fluorescence by FeXVII–FeXXIII is suppressed by resonant scattering (see Section 4). We add the annuli fluxes together taking into account Doppler blurring and gravitational redshift as described Chen, Halpern & Filippenko (1989). We have ignored the effects of light bending and our results should strictly be interpreted as time averaged since we do not take into account time dependence of the radiative transport, thermal transport, or illumination. A comparison of rapid X-ray variability ($\sim$ 1 ms; Rothschild et al. 1974), reflection region light crossing ($\sim 26r_g/c \sim 1$ ms), and plasma recombination/heating/cooling timescales ($\ll$ 1 ms) shows that at any instant disc plasma will be a 'quasi-thermal mess' state where electron temperature and ionization structure vary in a complicated way over the disc surface (note 'flickering' illumination need not remain fixed in one place). We also add the outward flux from the corona assuming it is an isotropic radiator.

Fig. 1 shows our results for discs at an inclination angle of 45 degrees. The total flux considered is roughly 77 per cent of the total flux from the disc plus corona system if the corona illumination is taken to extend to $\approx 26r_g$. Just above 7 keV a smeared edge feature is formed exactly as reported by Ebisawa (1991). Iron K$\alpha$ lines are conspicuous only in their absence; resonant Auger destruction destroys a large fraction of line photons and Doppler motions smear the rest from visibility. The low energy shape of the smeared edge is influenced by line photons which tend to fill it. The peaked structure around 0.7 keV in the $0.1L_{\rm Edd}$ case arises primarily from iron L lines (mainly from Fe XVIII–XXI), but there is also a contribution from the K line of O VIII. This structure may be compared with that shown in fig. 1 of Hess et al. (1994).

The contributions from some of the annuli in the disc operating at $0.1L_{\rm Edd}$ are shown in Fig. 2 (weighting by annulus area has been included). Several of the innermost zones produce a strong emission line from highly ionized iron, but this is blurred beyond recognition if $i \gtrsim 20$ degrees when Doppler and gravitational effects are included. The outer zones have little line emission due to resonant Auger destruction. When $i \lesssim 20$ degrees, the emission line from the innermost radii becomes apparent.

One important facet of our model is that an edge feature can be formed around 7.1 keV even from iron that is not cold, principally because of Doppler blurring, gravitational redshift and transverse Doppler shifts. The edge energy can vary between 7 and 8 keV depending primarily on the inclination and disc ionization parameter (its energy increases with both increasing inclination and increasing ionization parameter). This is consistent with current observations of the smeared edge (cf. sect 6.3 of Ebisawa 1991), and ASCA will allow a more precise test of this prediction of our model and reflection in BHC more generally.

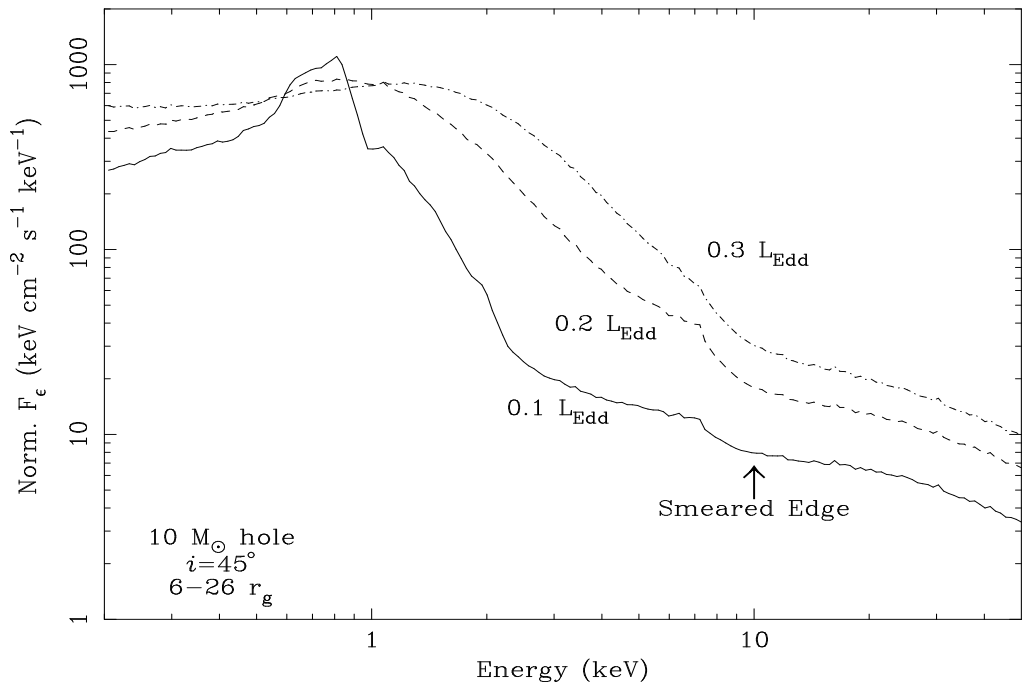

**Figure 1.** Spectra from the region 6–26$r_g$ of illuminated BHC discs at a 45 degree inclination angle around a 10 M$_\odot$ black hole. Note the prominent smeared edges and the lack of a visible iron K$\alpha$ line.

## 3 COMPARISON WITH OBSERVED SPECTRA

We have converted our $0.1L_{\rm Edd}$ model spectrum at $i = 45$ degrees into a fake X-ray spectrum assuming the response of one of the Gas Imaging Spectrometers, specifically GIS2, on ASCA (cf. Tanaka, Inoue & Holt 1994) and fitted the data over the 4–10 keV band with several models using the spectral fitting package XSPEC (Shafer et al. 1991). As expected a simple power-law is a very poor fit. To illustrate this we have fitted a power-law over the 4–6.8 keV band and extrapolated it to 10 keV. The ratio of the (fake) data to this model, shown in Fig. 3, clearly shows the beginning of the large broad iron edge and the lack of a line (the small deviations from the power-law below 6 keV are numerical artifacts of the discreteness of the annuli and the blurring processes). The maximum fractional depth, of about 23 per cent, is approximately twice as large as that observed in data of Cyg X-1 (e.g. Ebisawa 1991). A ten per cent increase in the direct continuum seen relative to the reflected spectrum, perhaps due to a different inclination angle than that assumed here, could easily reduce the predicted edge to agree with observations.

Ebisawa (1991) and Ebisawa et al. (1994) have modelled the absorption feature empirically with a 'smeared edge' model (denoted 'smedge' in XSPEC) in which the iron absorption cross-section, $\sigma$, above the edge energy $E_{\rm Edge}$ has the form $\sigma[1 - \exp(-(E - E_{\rm Edge})/E_{\rm W})]$. We confirm that this gives an excellent fit to the fake data from our models. At $0.1L_{\rm Edd}$ we find that a power-law with photon index 1.45, $E_{\rm Edge} = 6.93$ keV, $\tau_{\rm max} = 1.23$ and $E_{\rm W} = 3.72$ keV yields a fit in which the residuals are all less than 1.2 per cent of the data. $\tau_{\rm max}$ is the maximum optical depth in the smeared edge model. Ebisawa (1991) reports $E_{\rm Edge} \approx 7$ keV and $E_{\rm W} > 4$ keV from Ginga data of several BHC. At this stage we conclude therefore that for intermediate inclination angles the blurred absorption feature produced by our model has a similar shape to that observed from BHC, a success for the reflection model.

To assess the influence of resonant Auger destruction on the final iron K$\alpha$ line observed, we have constructed further fake data in which the disc in the range 6–26$r_g$ is assumed to emit the spectrum appropriate for the zone centred at $11r_g$, where the iron is principally FeXXV (Fig. 2) and Auger ionization cannot occur. A broad emission feature is then clearly seen at energies below the edge. Fitting these fake data with a model consisting of a power-law, a smeared edge and a Gaussian line yields an equivalent width for that line of 390 eV (the r.m.s. line width is 0.9 keV). This contrasts with the situation in our basic model at $0.1L_{\rm Edd}$ where no line is required (see Fig. 3); any line (which is no so broad as to encompass the edge) has an equivalent width less than 15 eV. This suggests that Auger destruction is an important reason why a detectable iron line is absent in our basic model. Doppler blurring at small radii is the other reason.

## 4 RESONANT AUGER DESTRUCTION

We now turn to an examination of the process of resonant Auger destruction of iron K$\alpha$ photons and its effects on the iron line emitted by a highly photoionized accretion disc. The detailed atomic physics of iron line formation is formidable as is illustrated by high resolution solar flare and tokamak spectra (e.g. Hill et al. 1979; Phillips et al. 1983; fig. 1 of Seely, Feldman & Safronova 1986; fig. 7 of Jacobs et al. 1989; fig. 5 of Beiersdorfer et al. 1993). We will of necessity restrict the scope of our treatment.

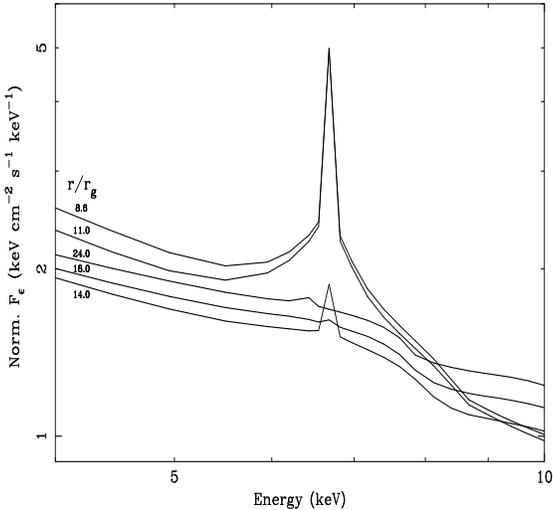

**Figure 2.** Contributions to the iron Kα line from annuli centered at the radii shown for a disc operating at $0.1L_{\rm Edd}$. The predominant iron ions at the surface of the disc as a function of radius are Fe XXV ($8.6r_g$), Fe XXV ($11.0r_g$), Fe XXV ($14.0r_g$), Fe XXI ($18.0r_g$) and Fe XIX ($24.0r_g$).

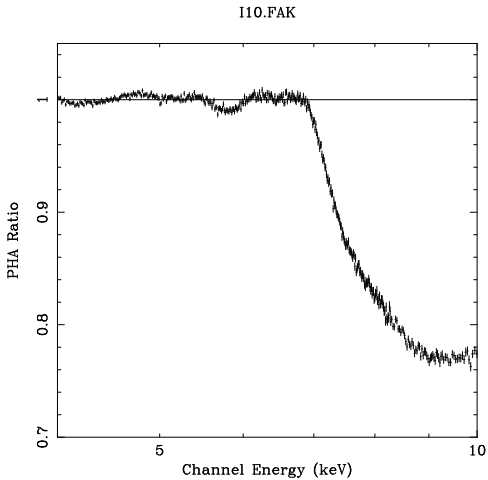

**Figure 3.** The data to model ratio when a power-law model is fit to faked ASCA GIS2 data (made from our $0.1L_{\rm Edd}$ computation viewed at 45 degrees) in the 4–6.8 keV band and then extrapolated to 10 keV. The beginning of the large broad iron K edge at $\sim 7$ keV and the lack of a line are apparent.

We consider iron line formation in an accretion disc plasma with $n_e \approx 2 \times 10^{21}$ cm$^{-3}$ and $T_e \approx 5 \times 10^6$ K. Many iron Kα photons produced by fluorescence of Fe XVII–Fe XXII will be destroyed by the combination of resonant trapping and the Auger effect. The Kα fluorescence lines for each of these iron species result from transitions to levels of the ground electronic configuration of the next more highly ionized species. This next more highly ionized species coexists with the fluorescing ion in the photoionized plasma, and collisional excitation establishes a statistical population among the various levels of its ground electronic configuration. Therefore these lines are strong resonance lines, and the Kα photons are easily destroyed via reabsorption in the line followed by autoionization (Auger effect).

To clarify the situation, consider the specific case of Kα fluorescence by carbon-like Fe XXI. A K-shell ionization of Fe XXI in its ground electronic configuration leaves the $1s2s^22p^2$ configuration of Fe XXII. Jacobs et al. (1989) have listed the eight most important radiative transitions from this excited configuration to the ground $1s^22s^22p$ configuration, along with calculated transition rates and branching ratios. Each transition has a spontaneous emission rate $A_{ul} \sim 10^{14}$ s$^{-1}$. assuming a Doppler line profile, the average opacity in one of these lines compared to the Thomson scattering opacity is

$$\frac{\bar\kappa_{\rm line}}{\kappa_{\rm T}} = \frac{n_l(\pi e^2/m_e c)f_{lu}}{n_e \sigma_{\rm T} \Delta\nu_{\rm D}}, \qquad (1)$$

where $n_l$ is the number density of ions in the lower level (either the $^2P_{1/2}$ or $^2P_{3/2}$ ground level of Fe XXII), $n_e$ is the number density of free electrons, $f_{lu}$ is the absorption oscillator strength, and $\Delta\nu_{\rm D}$ is the Doppler width. Adopting the abundances of Morrison & McCammon (1983), we get

$$\frac{\bar\kappa_{\rm line}}{\kappa_{\rm T}} \approx 660 \frac{g_u}{g_l}\left(\frac{n_l}{n_{\rm Fe}}\right)\left(\frac{A_{ul}}{10^{14}\,{\rm s}^{-1}}\right)\left(\frac{\lambda}{1.9\,{\rm Å}}\right)^3 \sqrt{\frac{10^6\,{\rm K}}{T}}, \qquad (2)$$

where $n_{\rm Fe}$ is the total number density of iron ions, $\lambda$ is the wavelength of the line, and $g_u$ and $g_l$ are the degeneracies of the upper and lower levels, respectively. Even if the local fraction, $n_l/n_{\rm Fe}$, of iron ions in the appropriate ground level is only a few per cent, the line opacity greatly exceeds the Thomson opacity. Since iron fluorescence occurs within a layer of Thomson depth $\tau_{\rm T} \sim 1$ (see George & Fabian 1991 and references therein), resonant trapping of the line photons is important. Following the original K-shell ionization or a resonant reabsorption, the probability of a radiative de-excitation (the 'fluorescence yield' $Y$) varies from a low of $Y \approx 0.1$ to a high of $Y \approx 0.8$, depending on the particular excited level involved. (Jacobs & Rozsnyai 1986 found an average fluorescence yield $\bar Y = 0.48$ for Fe XXI). Even for transitions with the highest fluorescence yields, only a few reabsorptions are needed to destroy most of the Kα line photons. A line photon is destroyed after diffusing only a fraction of a Thomson mean free path away from its original point of emission, and few line photons escape from the gas.

Fluorescence by beryllium-like Fe XXIII is inherently weak. Single-photon transitions would be expected to be unimportant for the $1s2s^2$ configuration following K-shell ionization. Although Chen et al. (1981) found that configuration interaction with $1s2p^2$ increases the fluorescence yield substantially, it is still only $Y = 0.11$. Resonant Auger destruction can reduce the yield even further if $2s \to 2p$ collisional excitations in the dense plasma produce a large population of Fe XXIV ions in the $1s^22p$ configuration (see Jacobs et al. 1989).

The fact that Auger destruction predicts a substan-tial weakening of K$\alpha$ lines from intermediate iron ionization stages suggests that it may be part of the solution to what would otherwise be an unresolved problem: Fe XVIII–Fe XXIV K$\alpha$ lines (which result from photoionization of Fe XVII–Fe XXIII) have never been observed from Thomson thick plasma in a cosmic X-ray source. (This is certainly the case for BHC, and we showed in Section 3 that Doppler blurring by itself has difficulty explaining the lack of iron lines in the reflection model for BHC spectra.) We encourage further research on the detailed treatment of iron K$\alpha$ transitions in the context of Thomson thick radiative transfer in order to address this important issue.

Another point worth noting is that bulk turbulent motions in the direction perpendicular to the disc plane might reduce the line opacities enough so that Fe XVIII–Fe XXIV line formation can take place. Bulk turbulent velocities larger than $c$ times the ratio of the line width ($\sim$ 1 eV) to the line energy would be required (about 40 km s$^{-1}$). Turbulent velocities of $\sim \alpha c_{\text{sound}}$ depend on the viscosity law chosen, and would be $\sim$ 30 km s$^{-1}$ if $c_{\text{sound}}^2 = P_{\text{gas}}/\rho$ (cf. sect. 7 of Pringle 1981; we use $\alpha = 0.1$ as per Sect. 2). Note in theory measurements of or upper limits on the Fe XVIII–Fe XXIV lines from BHC might be used to learn about the viscosity law in the inner parts of their discs.

## ACKNOWLEDGEMENTS

We gratefully acknowledge support from the College of the Holy Cross (RRR), the Royal Society (ACF) and the United States National Science Foundation and the British Overseas Research Studentship Programme (WNB). We acknowledge help from members of the Institute of Astronomy X-ray group and the creators of the XSPEC X-ray spectral fitting package. We acknowledge useful discussions with H. Mason, K. Phillips and Y. Tanaka.

## REFERENCES

Beiersdorfer P., Phillips T., Jacobs V.L., Hill K.W., Bitter M., von Goeler S., Kahn S.M., 1993, ApJ, 409, 846
Chen K., Halpern J.P., Filippenko A.V., 1989, ApJ, 339, 742
Chen M.H., Crasemann B., Karim K.R., Mark H., 1981, Phys Rev A, 24, 1845
Done C., Mulchaey J.S., Mushotzky R.F., Arnaud K.A., 1992, ApJ, 395, 275
Ebisawa K., 1991, PhD Thesis, Univ. of Tokyo
Ebisawa K. et al., 1994, PASJ, 46, 375
Galeev A.A., Rosner R., Vaiana G.S., 1979, ApJ, 229, 318
George I.M., Fabian A.C., 1991, MNRAS, 249, 352
Haardt F., Done C., Matt G., Fabian A.C., 1993, ApJ, 411, L95
Hess C.J., Paerels F., Kahn S.M., Liedahl D.A., Rogers R.D., 1994, in Holt S.S., Day C.S., eds, The Evolution of X-ray Binaries, AIP Press, New York, p. 166
Hill K.W. et al., 1979, Phys Rev A, 19, 1770
Inoue H., 1992, ISAS Research Note: Accretion Discs in Bright X-ray Binaries
Jacobs V.L., Rozsnyai B.F., 1986, Phys Rev A, 34, 216
Jacobs V.L., Doschek G.A., Seely J.F., Cowan R.D., 1989, Phys Rev A, 39, 2411
Kitamoto S., Takahashi K., Yamashita K., Tanaka Y., Nagase F., 1990, PASJ, 42, 85
Marshall F.E., Mushotzky R.F., Petre R., Serlemitsos, P.J., 1993, ApJ, 419, 301
Matt G., Perola G.C., Piro L., 1991, A&A. 247, 25
Matt G., Fabian A.C., Ross R.R., 1993, MNRAS, 262, 179
Morrison R., McCammon D., 1983, ApJ, 270, 119
Phillips K.J.H., Lemen J.R., Cowan R.D., Doschek G.A., Leibacher J.W., 1983, ApJ, 265, 1120
Pringle J.E., 1981, ARA&A, 19, 137
Ross R.R., 1978, PhD thesis, Univ. of Colorado
Ross R.R., Fabian A.C., 1993, MNRAS, 261, 74
Ross R.R., Fabian A.C., Mineshige S., 1992, MNRAS, 258, 189
Ross R.R., Weaver R., McCray R., 1978, ApJ, 219, 292
Rothschild R.E., Boldt E.A., Holt S.S., Serlemitsos P.J., 1974, ApJ, 189, L13
Seely J.F., Feldman U., Safronova U.I., 1986, ApJ, 304, 838
Shafer R.A., Haberl F., Arnaud K.A., Tennant A.F., 1991, XSPEC Users Guide. ESA Publications, Noordwijk
Shakura N.I., Sunyaev R.A., 1973, A&A, 24, 337
Tanaka Y., 1992, in Makino F., Nagase F., eds, Ginga Memorial Symposium. ISAS Press, Tokyo, p. 19
Tanaka Y., Inoue H., Holt S.S., 1994, PASJ, 46, L37
Ueda Y., Ebisawa K., Done C., 1994, PASJ, 46, 107